\documentclass[aps,prl,reprint,groupedaddress]{revtex4-1}

\usepackage{graphics,graphicx, array, afterpage}
\usepackage[labelformat=simple]{subcaption}
\usepackage{amsmath,amsfonts,amsthm,bm} 
\usepackage[export]{adjustbox}
\graphicspath{ {./figures/} }
\usepackage{hyperref}

\begin{document}
	
\newcommand{\ket}[1]{\left|#1\right>}
\newcommand{\bra}[1]{\left<#1\right|}
\newcommand{\braket}[2]{\left<#1|#2\right>}
\newcommand{\Ba}{$^{138}\mbox{Ba}^+$ }
\newcommand{\g}{$g^{(2)}(0)$\ }
\renewcommand\thesubfigure{(\alph{subfigure})}

\title{High Purity Single Photons Entangled with an Atomic Memory}

\author{C. Crocker}
\email[]{ccrocker@umd.edu}
\author{M. Lichtman}
\author{K. Sosnova}
\author{A. Carter}
\author{S. Scarano}
\author{C. Monroe}

\affiliation{Joint Quantum Institute, Center for Quantum Information and Computer Science, and Department of Physics,
	University of Maryland, College Park}

\date{\today}

\begin{abstract}
Trapped atomic ions are an ideal candidate for quantum network nodes, with long-lived identical qubit memories that can be locally entangled through their Coulomb interaction and remotely entangled through photonic channels.  The integrity of this photonic interface is generally reliant on purity of single photons produced by the quantum memory.  Here we demonstrate a single-photon source for quantum networking based on a trapped \Ba ion with a single photon purity of $ g^{2}(0)=(8.1\pm2.3)\times 10^{-5} $ without background subtraction.  We further optimize the tradeoff between the photonic generation rate and the memory-photon entanglement fidelity for the case of polarization photonic qubits by tailoring the spatial mode of the collected light.
\end{abstract}


\maketitle

Entanglement between flying photonic qubits and local memory qubits is an essential component of quantum communication networks and distributed quantum computers \cite{Cirac:97, Brown:16, Kimble:08, Sangouard:09, Ritter:12}.  
Trapped atomic ions provide a natural way to generate this entanglement, with pure and replicable quantum memories that can be locally entangled through their mutual Coulomb interaction \cite{Cirac:95, Molmer:99} and also emit nearly identical photons for networking. 
When photons are emitted from appropriate atomic excited states, the memory qubit can become entangled with the photonic qubit \cite{Duan:04, Monroe:14}.  
This entanglement is generally degraded if the atom is re-excited after a photon is emitted or background photons are present, thus the purity of the single-photon source is critical for high fidelity atom-photon entanglement \cite{Madsen:06, Maunz:07}.  
Moreover, for non-zero collection solid angles, the atomic radiation pattern does not perfectly map onto experimental polarization modes, limiting free-space entanglement fidelity in the case of polarization qubits \cite{Blinov:04}.  
Here, we demonstrate methods for reducing these errors by using different colors of light for excitation and collection, and by applying a custom aperture to maximize collected light while keeping polarization mixing errors low.

\begin{figure}
	\begin{subfigure}{.2\textwidth}
		\caption{}
		\includegraphics[width=\linewidth]{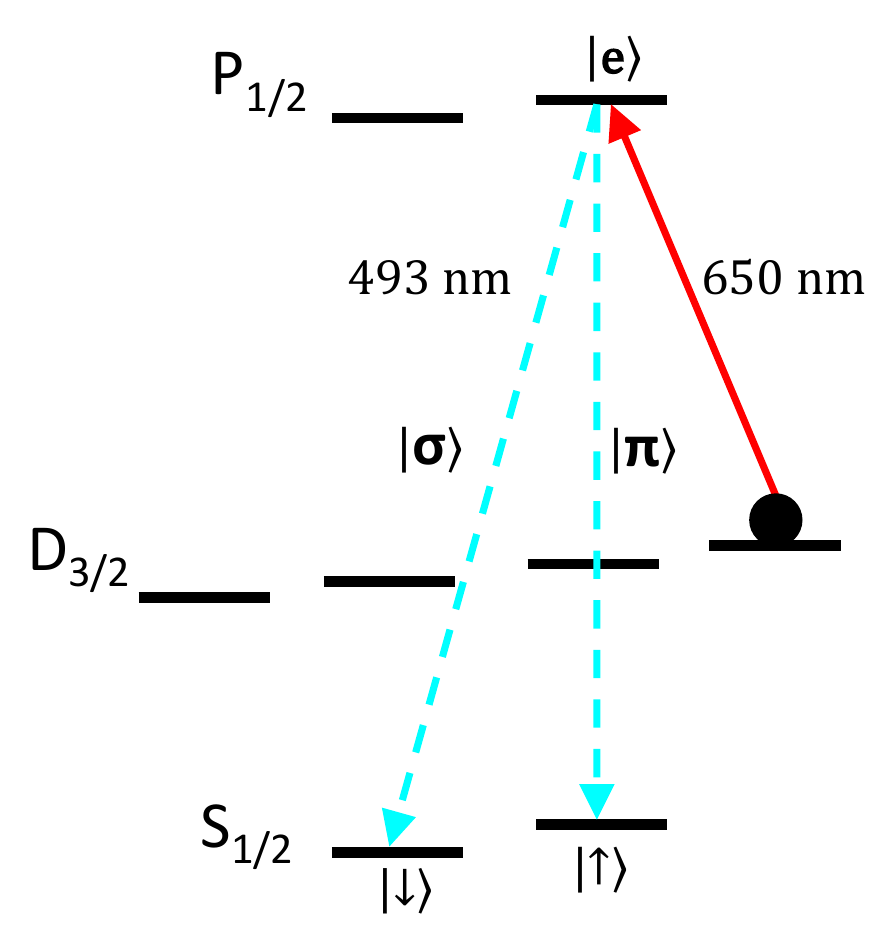}
		\label{fig:levelDiagram}
	\end{subfigure}
	\begin{subfigure}{.26\textwidth}
		\caption{}
		\includegraphics[width={\textwidth}]{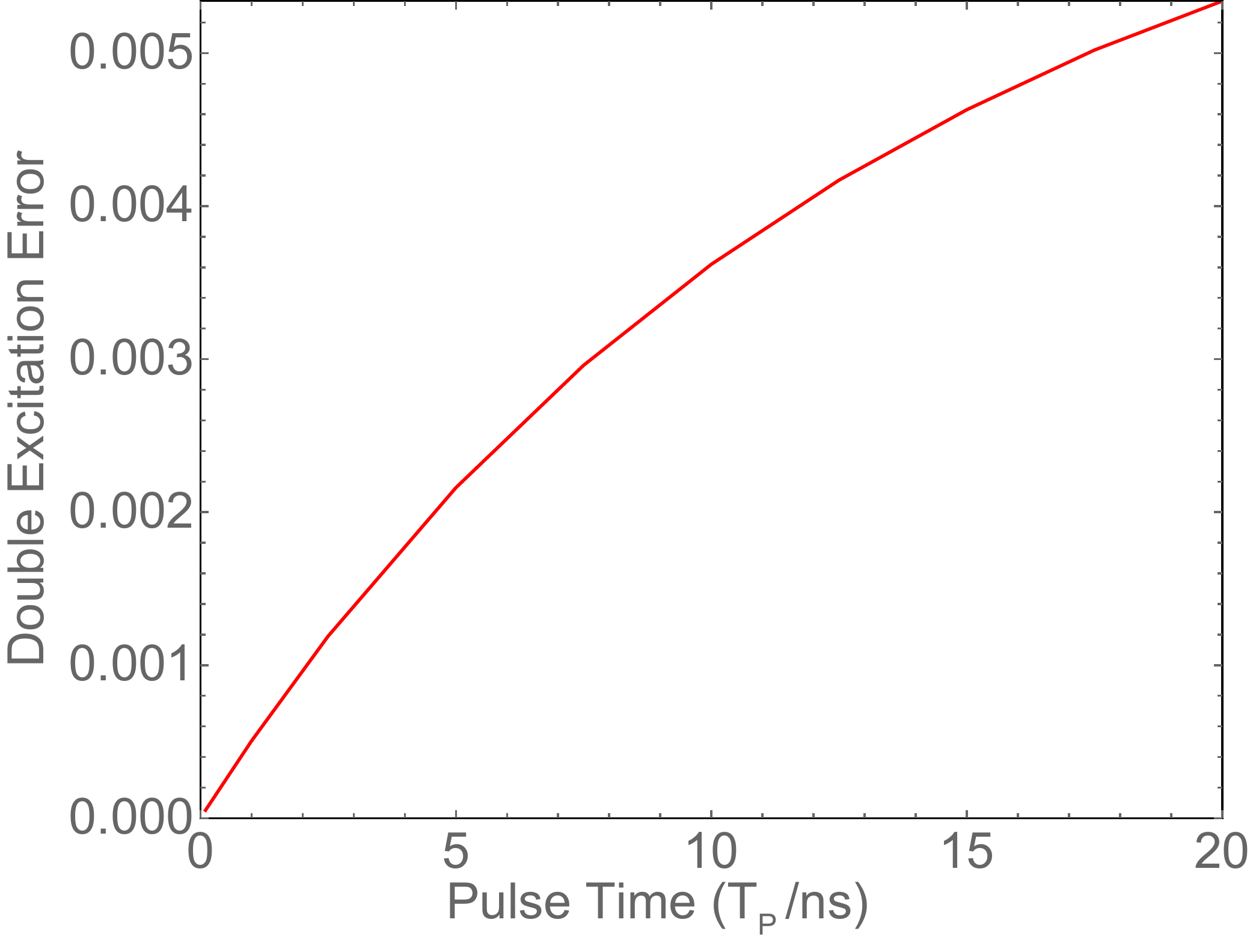}
		\label{fig:DoubleExcitationErrors}
	\end{subfigure}
	\begin{subfigure}{.3\textwidth}
		\includegraphics[width=\linewidth]{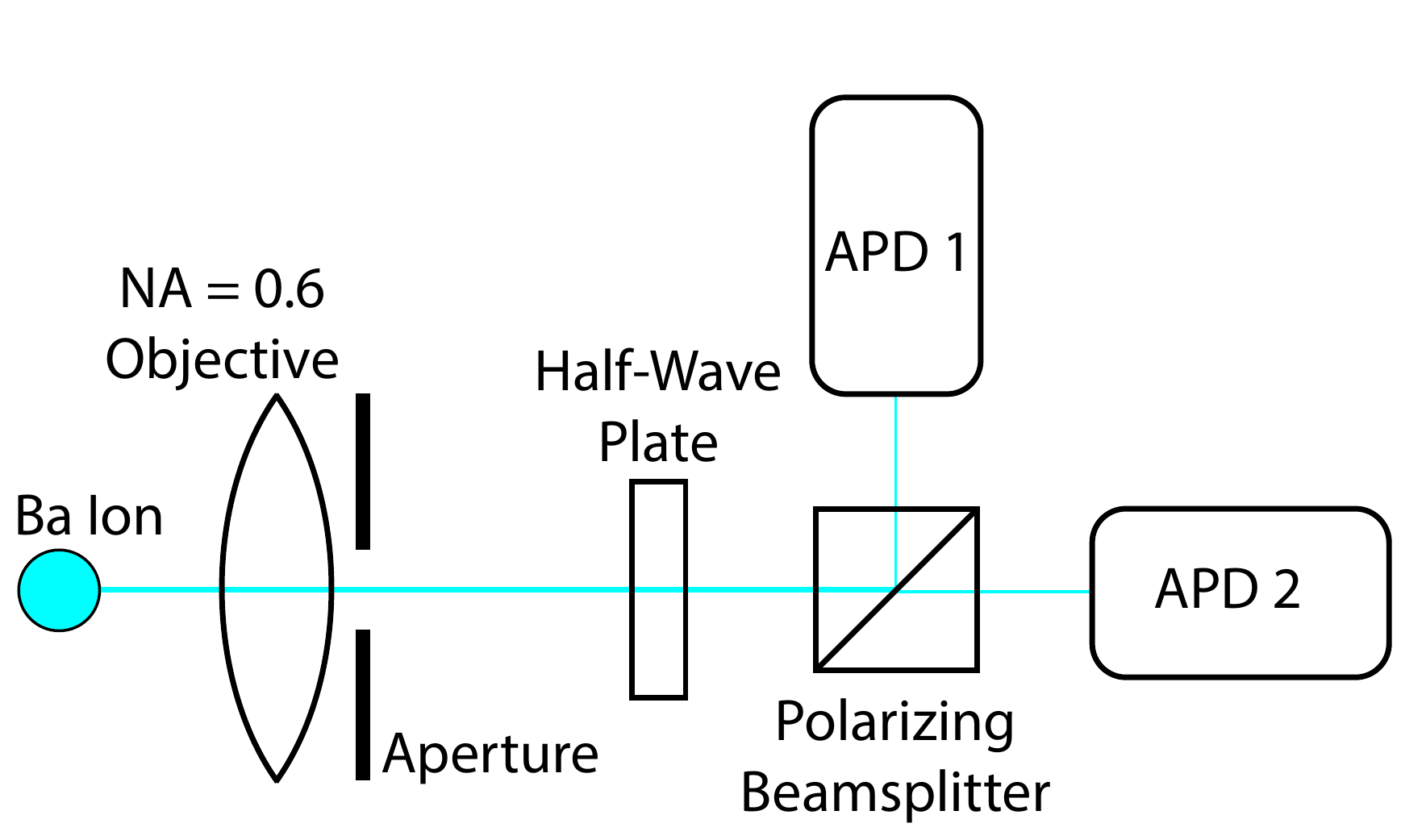}
		\caption{}
		\label{fig:polarizationAnalyzer}
	\end{subfigure}
	\caption{(a) Energy level diagram for \Ba atom.  (b)  Double excitation errors plotted as a function of pulse time $T_p$ assuming a Rabi rate of $\Omega=\pi/T_p$.  Note that even for pulses of order $\tau_e \sim 10$~nm the double excitation error is low. (c) Sketch of the setup used to collect light and analyze the polarization of photonic qubits.  Light is collected by a NA = 0.6 objective and then directed through a half-wave plate that can perform x-rotations on the polarization qubit.  Next is a polarizing, beam-splitting cube and a pair of APDs to detect the photon's polarization.  }
	\label{fig:levelsAndExcitations}
\end{figure}

The barium ion is an excellent candidate for trapped ion quantum network nodes \cite{Blatt:18, Blinov:14}.  
While most ions have their primary transitions in the UV wavelengths, barium has two lines in the visible range: a primary cooling transition at 493~nm ($6^2S_{1/2}$ to $6^2P_{1/2}$) and an auxiliary transition at 650~nm ($5^2D_{3/2}$ to $6^2P_{1/2}$). 
Compared to the UV transitions in most ions, photons in these visible wavelengths suffer less attenuation through optical fibers, permit access to a wide range of supporting photonic technologies, and can be converted to IR wavelengths for longer-distance networks  \cite{Siverns:17}.  
In this work we store the memory qubit in the two Zeeman levels of the \Ba $6^2S_{1/2}$ ground state: $ \ket{m_J=-1/2} \equiv \ket{\downarrow} $ and $ \ket{m_J=+1/2} \equiv \ket{\uparrow} $.  
To generate ion-photon entanglement, 493~nm photons are collected from decays from the $6^2P_{1/2}  \ket{J=1/2, m_J=+1/2} \equiv \ket{e}$ excited state (see Fig. \ref{fig:levelDiagram}) based on excitation from the $5^2D_{3/2} \ket{J=3/2, m_J= +3/2}$ state.

We first examine the effects of double excitations on the fidelity of ion-photon entanglement in this system. 
For probabilistic photon collection based on emission from an excited state $\ket{e}$, there are two mechanisms by which double excitations can introduce errors.  
In the first mechanism, the first photon emitted by the atom is collected, but the second photon is not.  
Here, the second excitation degrades the entanglement between the first photon and the state of the atom.  
In the second mechanism, the first photon is not collected but the second photon is.  
This situation still produces entanglement between the ion and the collected photon, but the scrambling of the atomic state after the first photon introduces errors into the fidelity of the desired entangled state.

Our previous work has shown ion-photon entanglement with \Ba by first pumping into $\ket{\downarrow}$ and exciting the atom to $\ket{e}$ with continuous-wave (CW) light at 493~nm \cite{Inlek:17}. 
Because this scheme uses the same line for excitation and collection light, it is susceptible to both types of double-excitation errors.  
These can be mitigated with a fast pulse of excitation light of duration $T_p \ll \tau_e$ where $\tau_e$ is the excited state lifetime \cite{Hucul:14}.  
Alternatively, the atom can be weakly excited with probability $P_e \ll 1$ such that the probability of double excitations scales as $P_e^{\ 2}$ \cite{Slodicka:13}.  
However, weak excitation reduces the overall entanglement success rate and forces a fundamental tradeoff between entanglement generation rate and fidelity \cite{Moehring:2007}. 

To avoid the difficulties caused by weak excitation and to eliminate the first mechanism of double excitation errors, we initialize the Ba$^+$ ion in the $5^2D_{3/2}$ manifold, excite with 650~nm light, and collect 493~nm fluorescence \cite{Siverns:17, Higginbottom:16, Bock:18}. 
Barium's $5^2D_{3/2}$ level features an 80~s lifetime \cite{Yu:97}, much longer than conceivable quantum operations, and its 3:1 branching ratio from the $6^2P_{1/2}$ state provides fast initialization and excitation.
Because the 650~nm excitation line is spectrally distant from the 493~nm collected photons, once a photon is collected there can be no further excitation events, eliminating the first mechanism for double excitations. 

We now estimate the expected error from the second mechanism of double excitation errors.
We evolve the optical Bloch equations for the excitation and emission in the regime where $T_p\sim\tau_e$, keeping track of whether the resulting $6^2S_{1/2}$ state population comes from decays from the desired $6^2P_{1/2}$ state ($\ket{e}$).
We find that favorable branching ratios and Clebsch-Gordan coefficients still lead to high-fidelity entanglement, even though $T_p \not\ll \tau_e$, as seen in Fig. \ref{fig:DoubleExcitationErrors}.
This significantly relaxes the need for ultrafast excitation pulses, since pulse durations $T_p\sim \tau_e = 10~\text{ns}$ can be created with a CW source and standard acousto-optic (AO) or electro-optic (EO) intensity modulators. 
The experiments presented in this paper are performed with 10~ns pulses generated by an AO modulator.

Next, we wish to demonstrate the efficiency of our system as a single photon source to verify a low level of double excitation errors \cite{Sosnova:18}.  
To show this, a \Ba ion is initialized in the the $5^2D_{3/2} (m_J=+3/2)$ stretch state by applying all polarizations of 493~nm light, and $\sigma^+$ and $\pi$ polarizations of 650~nm light.  
Next, a 10~ns pulse of $\sigma^-$ polarized light at 650~nm excites the atom to $\ket{e}$ (see Fig. \ref{fig:levelDiagram}).  
We collect the resulting 493~nm fluorescence photons with an $NA=0.6$ objective and through a polarization-analyzing 50/50 beamsplitter with avalanche photodiode (APD) detectors behind each port as shown in Fig. \ref{fig:polarizationAnalyzer}.  
To avoid collecting light from the initialization cycle the APDs are gated closed except for a 200~ns window triggered by the 650~nm pulse.

\begin{figure}
	\begin{subfigure}{.35\textwidth}
		\caption{}
		\includegraphics[width=1.0\linewidth]{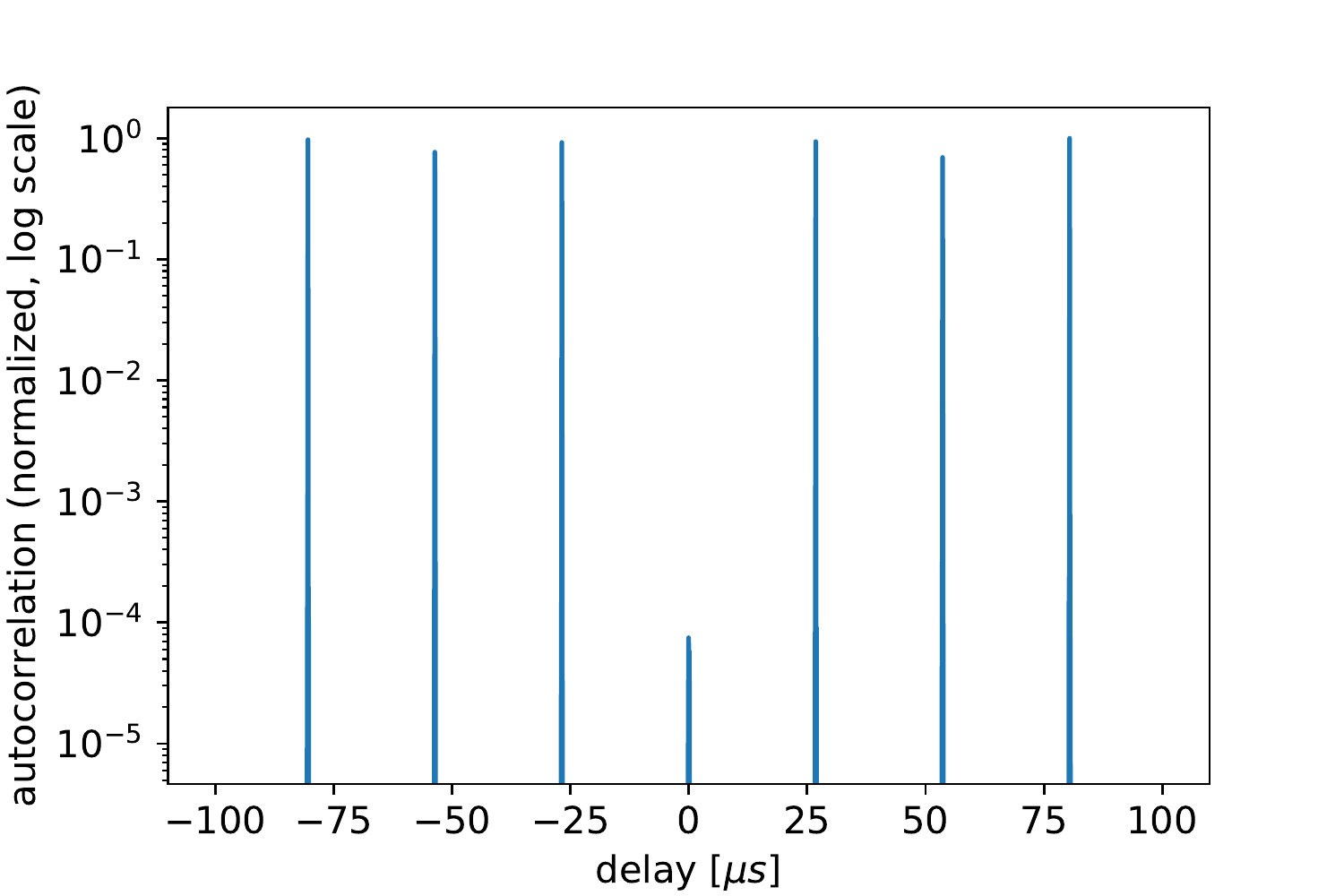}
		\label{fig:autocorrelation}
	\end{subfigure}
	\begin{subfigure}{.35\textwidth}
		\caption{}
		\includegraphics[width=1.0\linewidth]{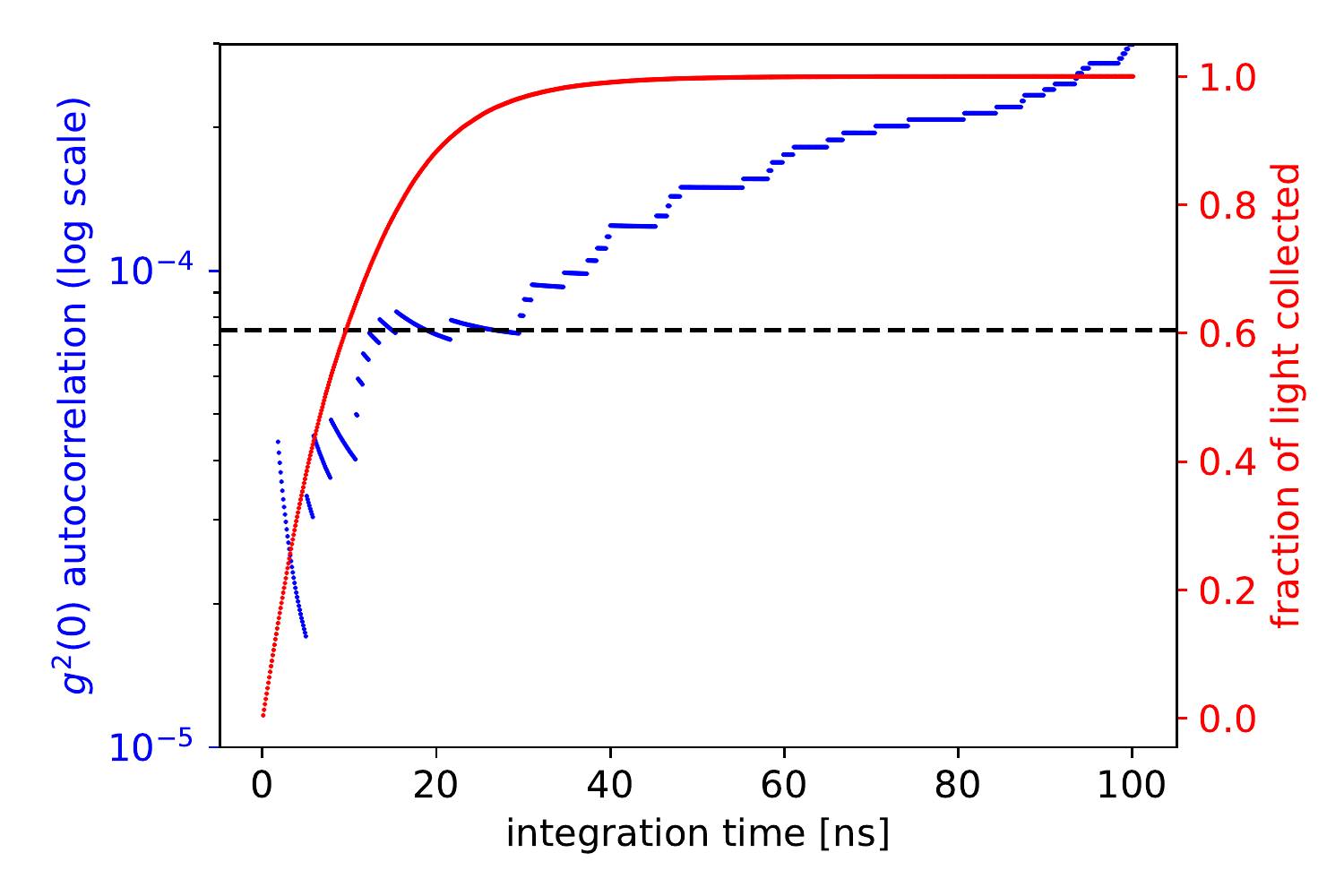}
		\vspace{-6px}
		\label{fig:g2time}
	\end{subfigure}
	\label{fig:g2}
	\caption{(a) Normalized second-order autocorrelation function.  26$\mu$s peak spacing corresponds to experimental repetition rate.  Strong suppression of $\tau=0$ peak demonstrates purity of single photon source.  (b) Calculated \g value (blue) and fraction of light collected (red) plotted as functions of integration time.  Dashed line represents the lowest reported \g value \cite{Schweickert:18}.}
\end{figure}

The normalized second-order autocorrelation function after integrating for 18 hours is plotted in Fig. \ref{fig:autocorrelation}. 
The strong suppression of the $\tau=0$ peak demonstrates the purity of the system as a single-photon source.  
In Fig. \ref{fig:g2time} we present $g^{(2)}(\tau=0) $ as a function of the integration window.  
We report a value of $g^{(2)}(0) = (8.1\pm2.3)\times10^{-5}$ using a 30~ns integration window.  
This window was chosen to provide a significant fraction of the collected photons as shown in the red curve of Fig. \ref{fig:g2time} while keeping dark counts from contaminating the signal. 
We measure 12(3) coincident events at zero delay ($\tau=0$) out of a total single photon count rate of 298,290(772), which is twice the size of adjacent peaks in Fig. \ref{fig:autocorrelation}.

This result represents the lowest value ever recorded for a source of indistinguishable photons \cite{Higginbottom:16} and is consistent with the lowest value reported in any system of $g^{(2)}(0) = (7.5\pm1.6)\times10^{-5}$ \cite{Schweickert:18}. 
Dark counts on our detectors limit \g to $\geq 3\times10^{-5}$, and we attribute the extra counts to observed transient light leakage through our AO modulators.  
This rate of multi-photon generation limits the contribution of the first mechanism of double excitation errors to a negligible value of $\leq4\times10^{-5}$.  
Moreover, as we show below, these single photons are entangled with the atomic memory, making them useful for networking applications.

Next we examine errors in ion-photon entanglement fidelity that can result from polarization mixing.  
There exist several different protocols for generating entanglement between an ion's spin state and the polarization of an emitted photon \cite{Duan:03, Simon:03, Moehring:07, Luo:09, Northup:14}.  
One common element is that they rely on faithfully mapping polarizations from atomic decays onto orthogonal polarization modes.  When the photons are collected in a single mode fiber, the polarization modes can be made orthogonal when the fiber mode is aligned along certain axes \cite{Kim:11, Lucas:fiberCoupling}.  But in free space, the polarization qubit becomes scrambled over finite solid angles of collection, leading to entanglement errors. 

\begin{figure}
	\centering
	\begin{subfigure}[m]{0.17\textwidth}
		\caption{}
		\includegraphics[width=\textwidth]{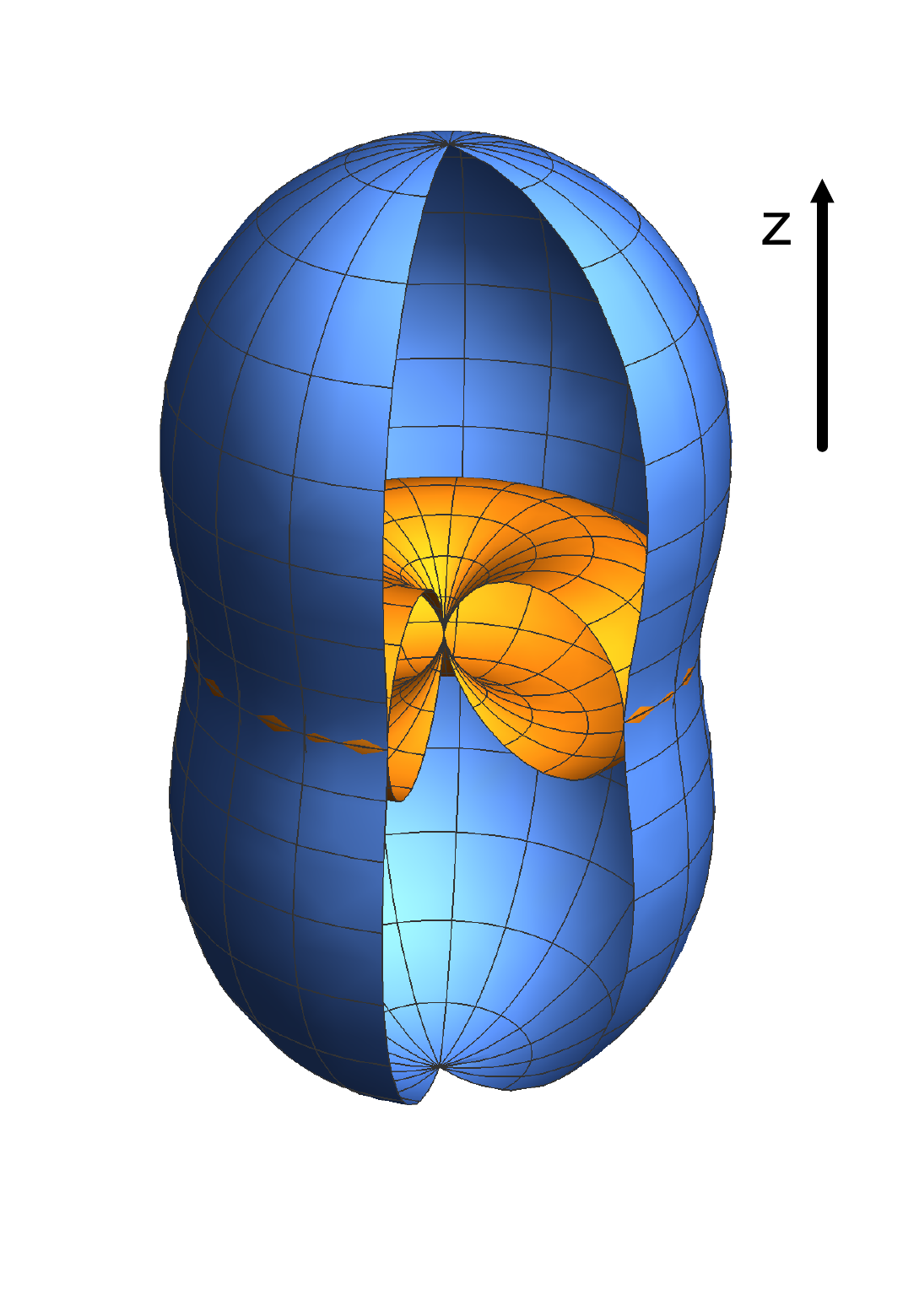}
		\label{fig:RadiationPatterns}
	\end{subfigure}
	~ 
	\begin{subfigure}[m]{0.17\textwidth}
		\caption{}
		\includegraphics[width=\textwidth]{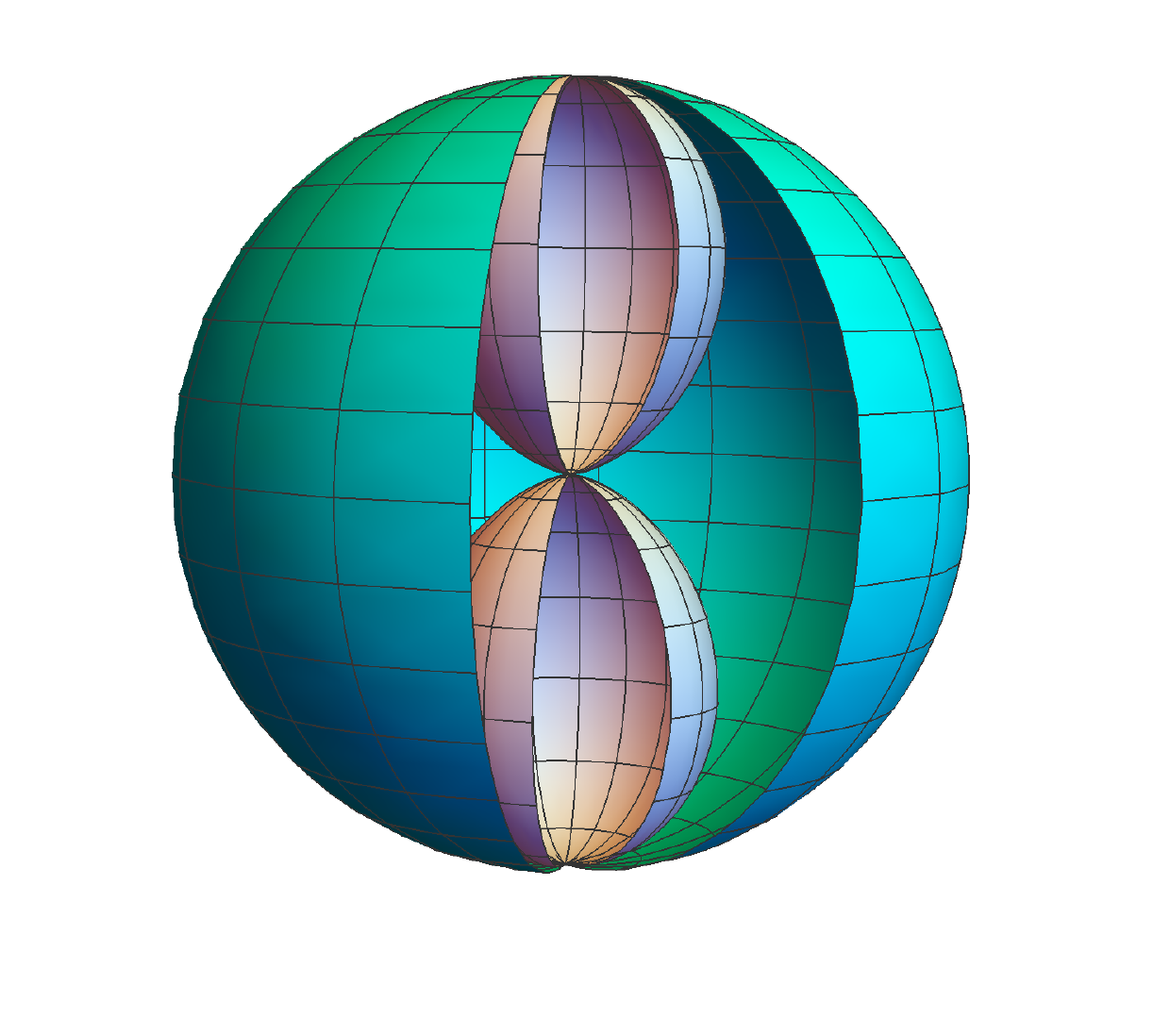}
		\label{fig:GoodBadSigma}
	\end{subfigure}
	\begin{subfigure}[m]{0.095\textwidth}
		\caption{}
		\includegraphics[width=\textwidth]{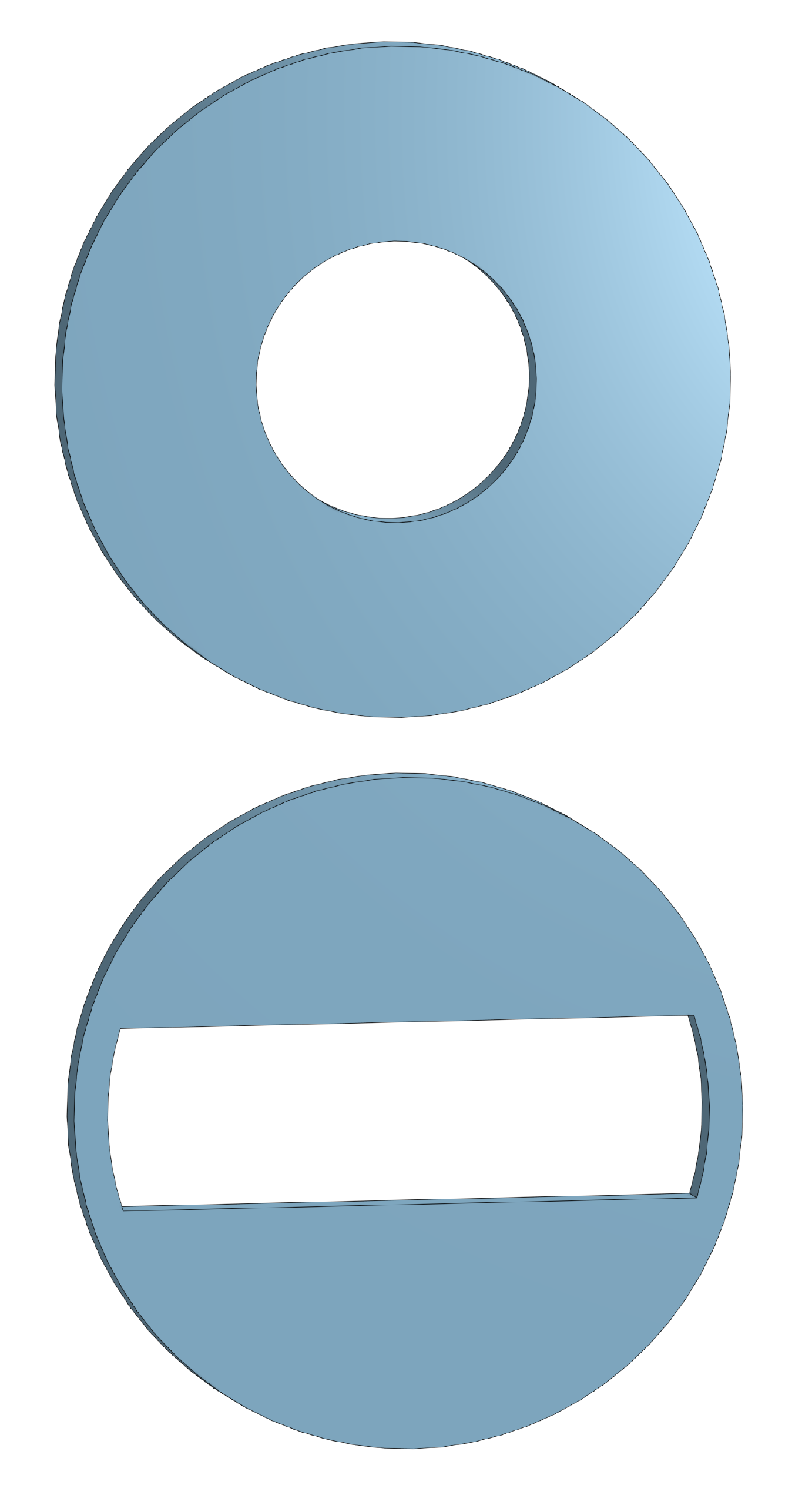}
		\label{fig:stops}
	\end{subfigure}
	~ 
	\caption{ (a) Spatial distribution of light from a $\sigma$-polarized (blue) and $\pi$-polarized (yellow) emission along the z quantization axis.  Note that in the x-y plane at polar angle $\theta=\pi/2$ there are equal amounts of $\sigma$ and $\pi$ emission.  (b) Decomposition of the $\sigma$ radiation pattern from (a) into horizontal (cyan) and vertical (light blue) linear polarization components.  At $\theta=\pi/2$ there is no vertical component.  (c) Two types of apertures are analyzed in this experiment.  Circular stop (top) used to restrict collection angle while maintaining a circular aperture.  Horizontal stop (bottom) used to restrict collection in the $\theta$ (vertical) direction while allowing full collection in the $\phi$ (horizontal) direction. }
	\label{fig:Polarizations}
\end{figure}

Consider a single atom with a quantization axis from an external magnetic field pointing in the z-direction undergoing spontaneous emission.  
The emitted photon can carry angular momentum of $\Delta m_z = +1, 0, \text{or} \ \nolinebreak{-1}$ quanta along z, and we will refer to these as $\sigma^+$, $\pi$, and $\sigma^-$ events respectively.  
The angular radiation patterns resulting from these decays are shown in Fig. \ref{fig:Polarizations}, $\bm{\pi} = i\sqrt{\frac{3}{8\pi}}\sin(\theta)\hat{\theta}$, and $\bm{\sigma^\pm} = ie^{\pm i\phi}\sqrt{\frac{3}{16\pi}}(\cos{\theta\hat{\theta}}\pm i\hat{\phi})$.

\begin{figure}
	\centering
	\begin{subfigure}[b]{0.22\textwidth}
		\includegraphics[width=\textwidth]{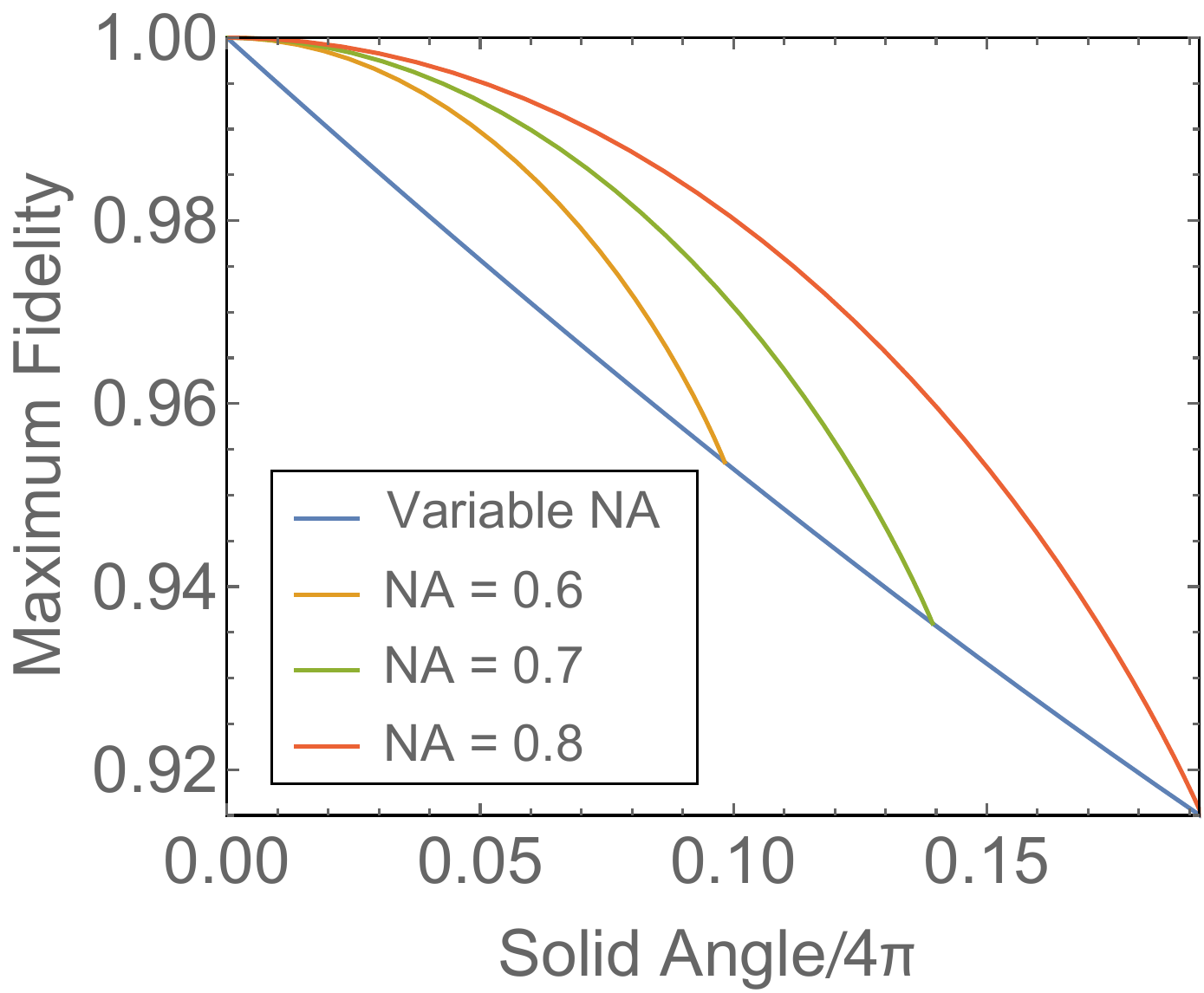}
		\caption{}
		\label{fig:wideSolidAngleErrors}
	\end{subfigure}
	~ 
	\begin{subfigure}[b]{0.23\textwidth}
		\includegraphics[width=\textwidth]{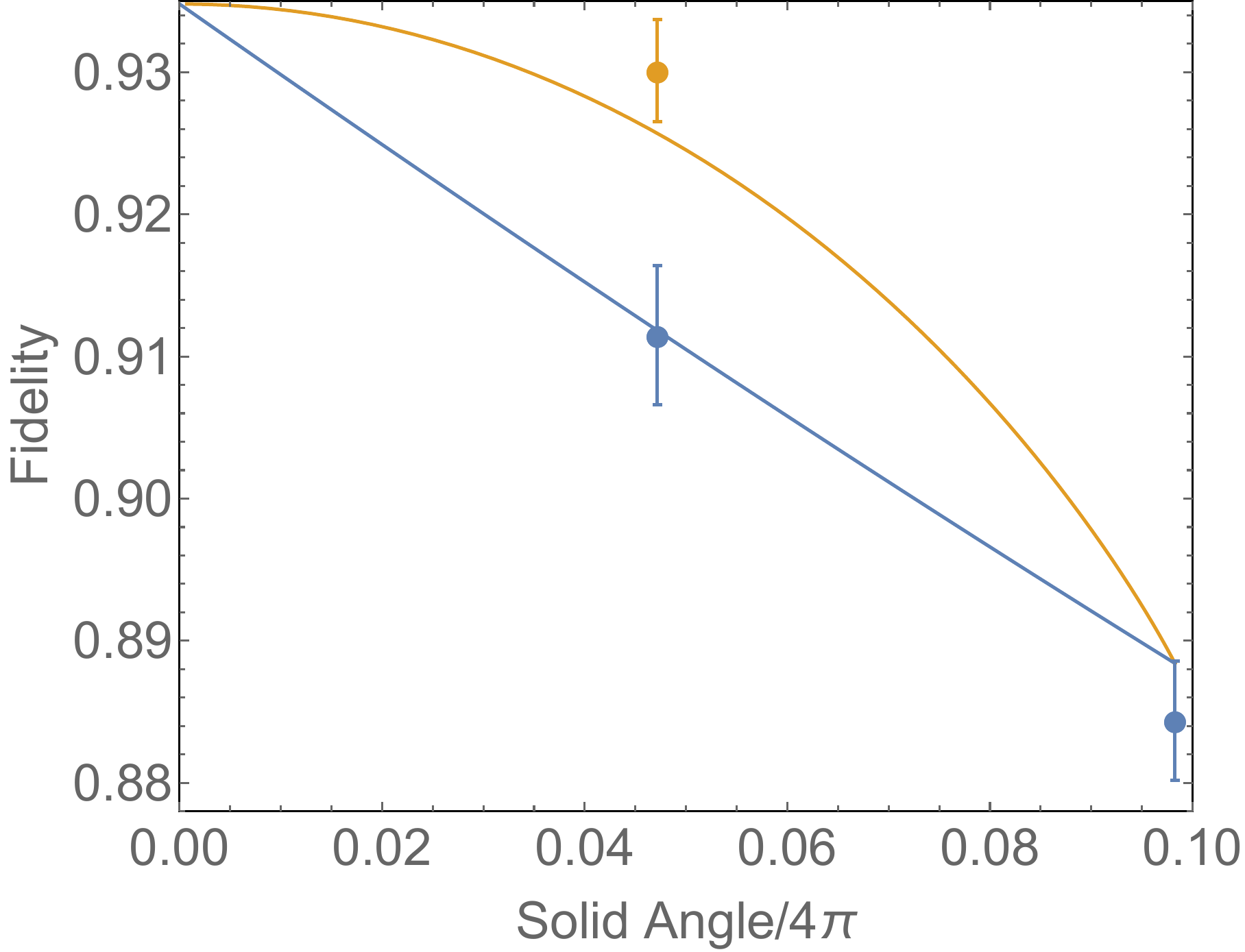}
		\caption{}
		\label{fig:polarizationMixingData}
	\end{subfigure}
	~ 
	\caption{(a) Theoretical scaling between solid angle of light collection and polarization-mixing errors on ion-photon fidelity.  The blue curve represents the scaling for a simple circular aperture.  The yellow, green, and red curves give the scaling assuming a fixed circular aperture of NA = 0.6, 0.7, or 0.8 respectively with added horizontal apertures that restrict collection in the $\theta$ direction (see bottom of Fig. \ref{fig:stops}). (b) The blue and yellow curves are the theoretical scaling curves from (a) applied to fidelity, including our other sources of error. The blue and yellow points show the data taken with the corresponding apertures applied. The error bars show 1$\sigma$ uncertainties.}
	\label{fig:PolarizationErrorTheory}
\end{figure}

We are interested in the case of light collected along the x-axis from a $P_{1/2} \rightarrow S_{1/2}$ transition such as the transition in \Ba shown in Fig. \ref{fig:levelDiagram}.  
After a photon is emitted the resultant atom-photon state is given by $\Psi_r = \frac{1}{\sqrt{2}}(\sqrt{P_\sigma}\ket{\downarrow\sigma} + \sqrt{P_\pi}\ket{\uparrow\pi})$ where $P_\sigma$ and $P_\pi$ are the probabilities of the collected photon coming from a $\sigma$ or $\pi$ decay.  
Here we are interested in mapping $\sigma^+$ (or $\sigma^-$) and $\bf{\pi}$ onto orthogonal, linear polarizations $\hat{H}=\hat{\phi}$ and $\hat{V}=\hat{\theta}$.  
This will create the desired maximally entangled state $\Psi_d = \frac{1}{\sqrt{2}}(\ket{\downarrow H} + \ket{\uparrow V})$.  
There are two potential sources of error.  First, as implied in Fig. \ref{fig:RadiationPatterns}, for large collection angles $P_\sigma \neq P_{\pi}$.  
Second, as shown in Fig. \ref{fig:GoodBadSigma}, $\sigma^+(\theta) = H(\theta)$ only at $\theta = \pi/2$.  
We calculate the resulting errors by first integrating the spatial distributions of $\sigma$ and $\pi$ decays to find $P_\sigma$ and $P_\pi$ as a function of the half-angle of our collection optics $\alpha_1$.  
Next we numerically integrate the $H$ and $V$ components of the $\sigma$ decays to find $P_{\sigma H}$ and $P_{\sigma V}$, the probabilities that a collected $\sigma$ photon is detected as $H$ or $V$.  
This traces over the spatial profiles of the photon and results in a mixed state $\rho_r$ with a diagonal term corresponding to $\ket{\downarrow V}$ population with magnitude $P_{\sigma V}$.
We then define the error $\epsilon=1-F=1-\bra{\Psi_d}\rho_r\ket{\Psi_d}$ where $F$ is the fidelity of the ion-photon entanglement.  
This error is plotted as a function of solid angle collected in the blue curve of Fig. \ref{fig:wideSolidAngleErrors}. 
This result confirms that as larger solid angles are used to improve entanglement generation rates, the fidelity of free-space ion-photon entanglement will suffer.  

Both types of errors shown here increase for light emitted further off-axis (away from $\theta = \pi/2$), but are independent of $\phi$.  
This suggests that it may be possible to achieve a more favorable trade-off between rate and fidelity by blocking light in the $\theta$ direction.  
To analyze this possibility, we consider light collected by a lens with collection half-angle $\alpha_1$ with a set of stops that limit collection in the vertical direction to the range $\theta \in [\pi/2 - \alpha_2, \pi/2 + \alpha_2]$ where $\alpha_2 \leq \alpha_1$ (Fig. \ref{fig:stops}). 
The error calculations are then repeated and the results are plotted in Fig. \ref{fig:wideSolidAngleErrors} as a function of total solid angle that passes through the apertures for a variety of values of $\alpha_1$.  
The results confirm that the horizontal stops provide a favorable trade-off between light collection and fidelity for free-space ion-photon entanglement.

To experimentally examine this polarization mixing, we perform ion-photon entanglement using a single trapped \Ba atom.  
First, we use 650~nm excitation to generate an entangled ion-photon pair as described in previous sections. 
For this experiment, we also make use of the half-wave plate that can rotate the photon's polarization before the polarization measurement. 
Additionally, ion spin state rotations and readout are performed using the methods described in \cite{Inlek:17}.  

\begin{figure}
	\centering
	\begin{subfigure}[b]{0.23\textwidth}
		\includegraphics[width=\textwidth]{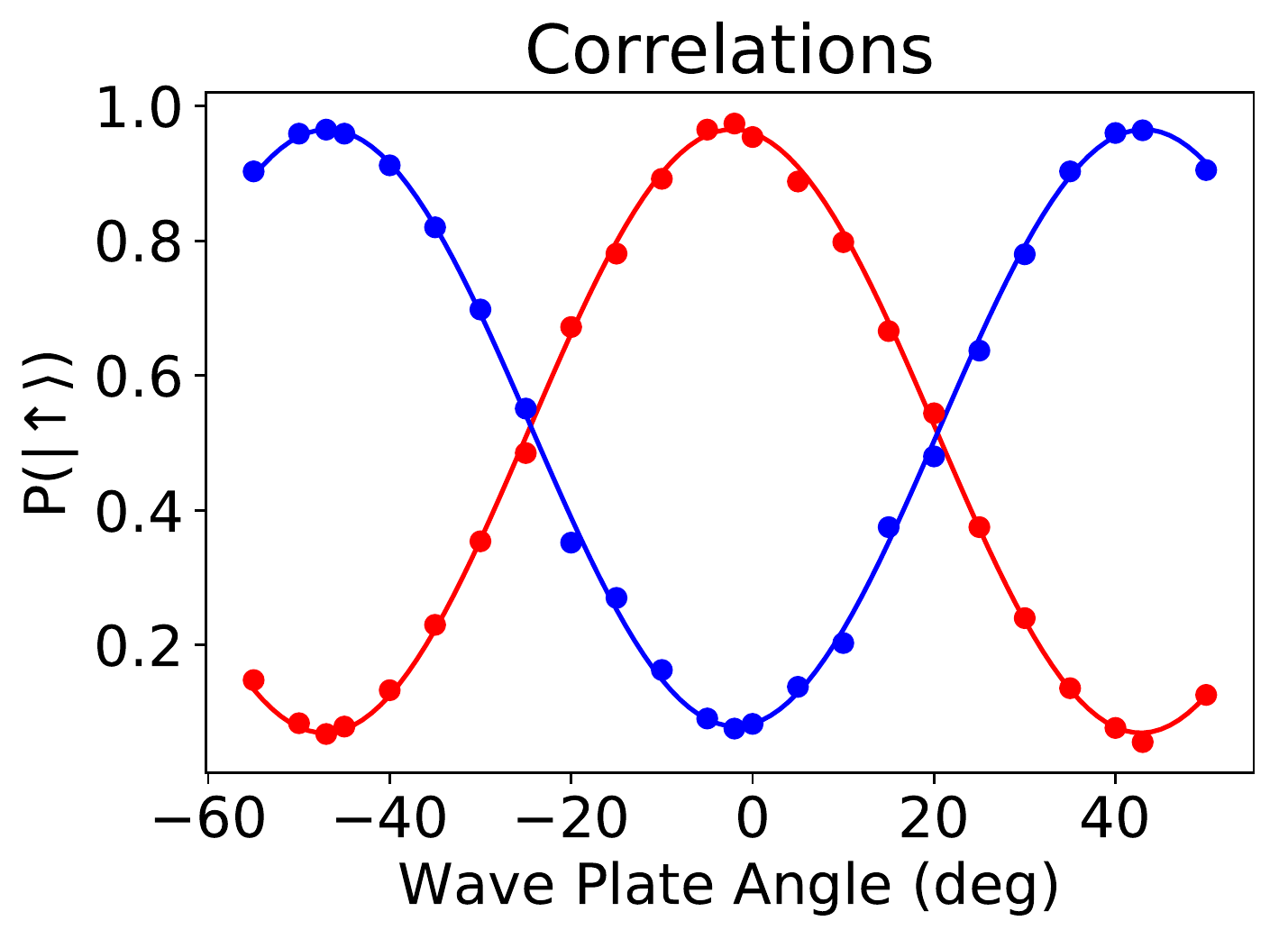}
		\caption{z-basis correlations}
		\label{fig:correlations}
	\end{subfigure}
	~ 
	\begin{subfigure}[b]{0.23\textwidth}
		\includegraphics[width=\textwidth]{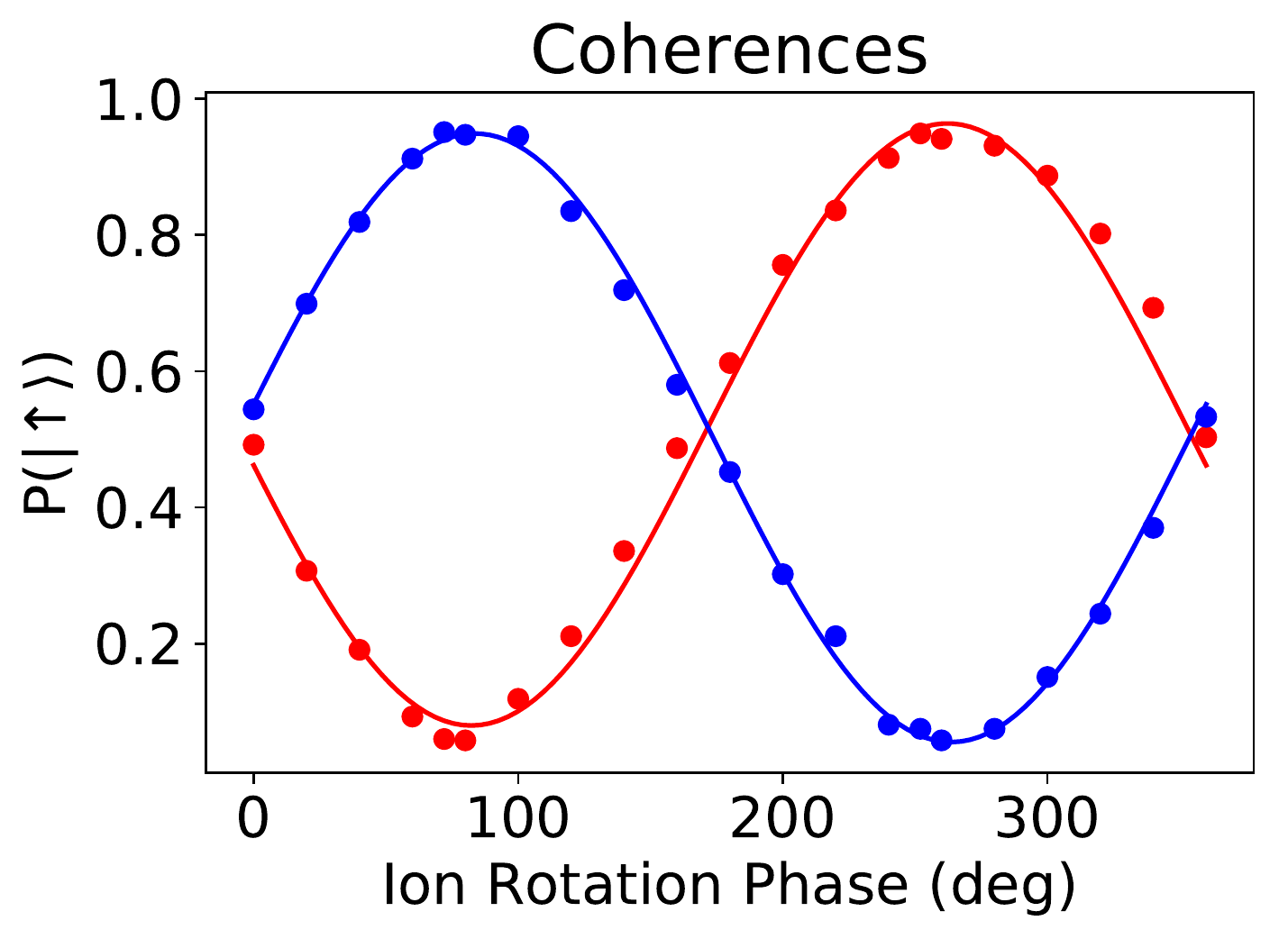}
		\caption{x-basis correlations}
		\label{fig:coherences}
	\end{subfigure}
	~ 
	\caption{ (a) Ion-photon correlation results as a function of wave plate rotation angle.  The red (blue) curve shows the probability of finding the ion in the $\ket{\uparrow}$ state when the photon is detected on APD1 (APD2). No stops were used for these experiments. (b) Coherences in the y-basis are taken by setting the half-wave plate to perform a $\pi/2$ rotation on the photon and then applying a $\pi/2$ pulse on the ion with a varying phase. }
	\label{fig:Entanglement}
\end{figure}

To demonstrate entanglement, we first directly show correlations between the state of the ion and the photon by analyzing these correlations as a function of photon rotation angle. 
Next, the coherences are measured by fixing the wave plate angle to rotate the polarization by $\pi/2$ and performing a $\pi/2$ Raman rotation on the atom with a variable phase.  
These results are plotted in Fig. \ref{fig:Entanglement} and show an ion-photon entanglement fidelity of $F=0.884(4)$ when light is collected over the entire circular 0.6 NA of the lens. 
Intrinsic polarization mixing for this size of aperture accounts for a fidelity loss of 0.046; we attribute the remaining errors to imperfect state initialization and readout, intensity and phase noise on the Raman beams used to analyze the coherences, and polarization mixing in the collection optics \cite{Hucul:14}.  
The analysis from previous sections indicates that errors from double excitations contribute an error of $<0.004$.

To analyze the effects of spatial filtering on ion light, various optical stops were inserted immediately after the microscope objective (see Fig. \ref{fig:stops}).  
These apertures were designed to block half of the solid angle either symmetrically (circular stops) or in only the $\theta$ direction (horizontal stops).  
After inserting the stops, the entanglement experiments were repeated; the circular stops gave a fidelity of 0.912(5) and the horizontal stops improved this further to 0.930(4).  
These results are plotted along with the theory curves in Fig. \ref{fig:polarizationMixingData} and confirm that, by taking into consideration the spatial profile of the atomic fluorescence, we can maximize fidelity gained by sacrificing rate.

For future quantum networks, the pure entangled photons demonstrated in this letter can be combined with techniques for performing local operations described in previous works \cite{Inlek:17} to construct a modular node consisting of a superior Yb memory qubit and visible photon flying qubits.  
This Yb memory is unaffected by the photon generation process, allowing for local operations or storage while the Ba-photon link is created.  
Multiple nodes can be connected together using a photonic Bell state analyzer \cite{Maunz:07} to create a distributed network for quantum information processing\ \cite{Monroe:14}.

\begin{acknowledgments}
We acknowledge the useful discussions with D. Nadlinger and D. Lucas.  
This work was supported by the ARO with funds from the IARPA LogiQ program and the MURI on Modular Quantum Systems, the AFOSR project on Quantum Networks, the ARL Center for Distributed Quantum Information, and the National Science Foundation Physics Frontier Center at JQI.
\end{acknowledgments}

\end{document}